# Suppression of Antiferromagnetism in Electron-Doped Cuprate $T'$-La$_{2-x}$Ce$_x$CuO$_{4\pm\delta}$

C. Y. Tang,[1,2,*] Z. F. Lin,[1,2,*] J. X. Zhang,[3] X. C. Guo,[1,2] J. Y. Guan,[1,2] S. Y. Gao,[1,2] Z. C. Rao,[1,2] J. Zhao,[1,2] Y. B. Huang,[4] T. Qian,[1,5] Z. Y. Weng,[3] K. Jin,[1,5,†] Y. J. Sun,[1,5,†] and H. Ding[1,2,5,†]

[1] *Beijing National Laboratory for Condensed Matter Physics and Institute of Physics, Chinese Academy of Sciences, Beijing 100190, China*
[2] *University of Chinese Academy of Sciences, Beijing 100049, China*
[3] *Institute for Advanced Study, Tsinghua University, Beijing 100084, China*
[4] *Shanghai Advanced Research Institute, Chinese Academy of Sciences, Shanghai 201204, China*
[5] *Songshan Lake Materials Laboratory, Dongguan, Guangdong 523808, China*

**Abstract**
We performed systematic angle-resolved photoemission spectroscopy measurements *in-situ* on $T'$-La$_{2-x}$Ce$_x$CuO$_{4\pm\delta}$ (LCCO) thin films over the extended doping range prepared by the refined ozone/vacuum annealing method. Electron doping level ($n$), estimated from the measured Fermi surface volume, varies from 0.05 to 0.23, which covers the whole superconducting dome. We observed an absence of the insulating behavior around $n \sim 0.05$ and the Fermi surface reconstruction shifted to $n \sim 0.11$ in LCCO compared to that of other electron-doped cuprates at around 0.15, suggesting that antiferromagnetism is strongly suppressed in this material. The possible explanation may lie in the enhanced $-t'/t$ in LCCO for the largest La$^{3+}$ ionic radius among all the Lanthanide elements.

In the pursuit of the grand understanding of high-temperature cuprate superconductivity, people often find the electron-doped ($n$-type) cuprates puzzling and somewhat awkward, mainly owing to a much more robust antiferromagnetic (AF) phase and the absence of pseudogap phase in the $n$-type cuprates [1,2]. It is not clear up to now whether the "electron-hole" asymmetry are fundamental (i.e. due to different residing orbitals [1] or different correlation strengths caused by electron or hole charges [3,4]) or accidental (i.e. due to a band effect caused by the structure difference in the $n$-type cuprates [5,6]). The general formula unit of electron-doped cuprates is usually given as Ln$_{2-x}$Ce$_x$CuO$_{4\pm\delta}$ where Ln = La, Pr, Nd, Sm, Gd with gradually decreasing ionic radius. They are the so-called $T'$ families due to the lack of apical oxygen ions, and as-grown samples become superconducting only after proper reduction process, which is speculated to remove apical or planar oxygen ions, or compensate Cu vacancies [7]. Among the Ln$_{2-x}$Ce$_x$CuO$_{4\pm\delta}$ systems, La$_{2-x}$Ce$_x$CuO$_{4\pm\delta}$ (LCCO) has the highest $T_c$ and the narrowest AF region [2,8], which may be in the closest proximity to their hole-doped counterpart La$_{2-x}$Sr$_x$CuO$_{4\pm\delta}$ (LSCO). However, the $T'$-LCCO can only be stabilized in the thin-film form, and underdoped film is difficult to grow in the single $T'$ phase [7,9], thus there lack systematic experimental measurements, including angle-resolved photoemission spectroscopy (ARPES) results [10], for this material.

In this paper, we adopted the two-step ozone/vacuum annealing procedure to modify the electron doping of LCCO films [2], and performed *in-situ* ARPES experiments on these films. High-quality LCCO films were grown by pulsed laser deposition (PLD) with Ce concentrations $x = 0.1, 0.19$ and thickness of 150–200 nm. During the first annealing step, films were annealed at 700 °C and the ozone partial pressure of $5 \times 10^{-7}$ torr for 30 mins in an ozone-assisted molecule beam evaporation (MBE) system. This high-temperature ozone annealing restores freshness of sample surfaces by removing the impurities absorbed in the air, and also reduces the electron doping level by adding more oxygen ions into the sample. The films were subsequently annealed under high vacuum at different temperatures between 400 °C and 700 °C for 30 mins to remove excess oxygen ions, and therefore their electron doping levels can be continuously changed and their electronic structures can be measured by ARPES *in-situ*. The electron doping level $n$ can be estimated from the measured Fermi surface (FS) volume by the Luttinger sum rule [11–14]. Using this method, we were able to measure ARPES spectra on $La_{2-x}Ce_xCuO_{4\pm\delta}$ films from $n \sim 0.05$ to 0.23 across the whole superconducting dome with $x = 0.1$ and 0.19 samples. This refined ozone/vacuum annealing method gives us a good opportunity to explore the evolution of FS and band structure with doping in LCCO.

For $n \sim 0.05$–0.19, ARPES measurements were performed with a VG Scienta R4000 analyzer and a VUV helium plasma discharge lamp in our laboratory at the Institute of Physics. He IIα (40.8 eV) photons were used and the base pressure of the ARPES system is $4 \times 10^{-11}$ torr. All measurements were carried out at 30 K without special statement. For $n \sim 0.23$, ARPES was measured at the Dreamline beamline of the Shanghai Synchrotron Radiation Facility (SSRF) with a Scienta Omicron DA30L analyzer at ~ 20 K, with the photon energy of 55 eV.

Fig. 1(a) is the phase diagram of LCCO from transport and angular magnetoresistance measurements [8,15,16], the blue marks represent the doping levels of films after ozone/vacuum annealing, with the Ce concentrations $x$ corresponding to 0.1 and 0.19. Fig. 1(b) shows the reflection high-energy electron diffraction (RHEED) patterns of a pristine film and that after ozone/vacuum annealing. The surfaces of *ex-situ* LCCO films are contaminated by air, which can be seen from the weak RHEED pattern. After the two-step ozone/vacuum annealing process, much clearer RHEED patterns and the appearance of Kikuchi lines indicate the recovery of freshness of sample surfaces, on which ARPES measurements can be performed afterwards.

The general electronic structure at a representative doping level $n \sim 0.06$ is illustrated in Fig. 2, where the spectra weight is suppressed at the cross point of FS and AF zone boundary, and the cross point is recognized as hotspot. Fig. 2(a) is the band dispersion measured at 10 K, and cut 1 to cut 4 are referred to as node, hotspot, near hotspot and antinode, respectively, as indicated in Fig. 2(b). The absent spectral weight at the upper half of FS is due to the matrix element effect. The band dispersion at the hotspot displays a large gap induced by the $(\pi, \pi)$ AF scattering, and near the hotspot has a noticeable kink at the AF zone boundary. Fig. 2(c) is the half width at half maximum (HWHM) of the momentum distribution curves (MDCs) extracted from these cuts. The overall MDC widths are relatively small, suggesting that the films are high quality after the ozone/vacuum annealing. The MDC width at the antinode is much wider than the one at the node, suggesting that the node has more well-defined quasiparticles while the antinode suffers from an enhanced scattering rate, similar to hole-doped cuprates [17,18].

The good quality of data paves a new platform for studying the doping dependence of the electronic structure in LCCO. We next turn attention to the evolution of FS. Fermi surfaces from $n \sim 0.05$–$0.23$ are displayed in Figs. 3(a1)–(a7). The trend of the tight-binding (TB) fitting from highly underdoped to highly overdoped film can be seen from the black curve to the red curve in Fig. 3(b). At $n \sim 0.05$, ARPES spectrum shows a metallic behavior with no charging gap, while transport and low-energy muon spin rotation (LE-μSR) measurements suggest an insulating AF behavior at 0.06 [8,9,19]. The long-range AF order is suppressed in ozone/vacuum annealed LCCO films compared to as-grown samples, possibly resulting from higher efficiency of reducing apical oxygens by the annealing process. While the Fermi surface reconstruction (FSR) at the underdoped region of LCCO is similar to other electron-doped cuprates, it should be noted that the FSR is observed only below $n \sim 0.11$, where no spectral weight is observed at the hotspot by integrating over $E_F \pm 15$ meV. Moreover, the disappearance of FSR happens at around 0.11 at the measured temperature 30 K is consistent with resistivity and Hall results [16]. This is much lower than that in other electron-doped cuprates at about 0.15 [13,20–22]. ARPES measurements at electron doping levels $n \sim 0.14$–$0.23$ were performed on $x = 0.19$ film, all the spectra show similar behavior with full circular FSs. Besides the fitting of FSs, we also display the nodal and antinodal band dispersions at $n \sim 0.23$ in Figs. 3(c)–3(d). By fixing the scaled chemical potential $\mu/t$ and scaled next-nearest-neighbor hopping $-t'/t$ as for the FS at $n \sim 0.23$, the TB fitting results with the nearest-neighbor hopping term $t = 0.3$ agree well with the experimental band dispersions. The well-fitted bands further support the accuracy of the estimated doping levels.

To get a clearer vision of the suppression of AF, we focus on the evolution of band dispersions at the hotspot from $n \sim 0.05$ to $0.14$ in Figs. 4(a)–4(e). The EDCs at $k_F$ are summarized in Fig. 4(f). The AF gap gradually closes as $n$ increases, and the maximum leading-edge (LE) shift is about 25 meV. But the LE gap is around 50 meV in PLCCO at $n \sim 0.045$ and a quantitative AF gap at the hotspot was estimated to be about 80 meV in NCCO at $x = 0.15$ [13,20,21]. There is a possible explanation for the weakening of AF in LCCO. By assuming the next-next-nearest-neighbor Cu hopping $t'' = 0$ in the TB model, the ratio $-t'/t$ generally gets larger with increasing doping levels, which can be seen from the increasing curvature of FS as shown in Fig. 3(b). While $-t'/t = 0.40, 0.23$, and $0.21$ for $Nd_{1.85}Ce_{0.15}CuO_4$ (NCCO), $Sm_{1.85}Ce_{0.15}CuO_4$ (SCCO) and $Eu_{1.85}Ce_{0.15}CuO_4$ (ECCO), respectively, $-t'/t = 0.46$ for LCCO near the optimal doping $n \sim 0.11$, as shown in Fig. 4(g). Note that by assuming $-t''/t' = 0.5$, $-t'/t = 0.23, 0.2, 0.12, 0.11$ for LCCO, NCCO, SCCO and ECCO, respectively [23]. The similar tendency demonstrates that with larger $Ln^{3+}$ ionic radius, $-t'/t$ becomes larger. It was suggested by previous optical and Raman experiments that the charge transfer gap $\Delta$ and the AF exchange interaction $J$ decrease with increasing Cu-O distance [24,25]. Our work, from the aspect of band structure, further supports that AF is likely suppressed by the enhanced next-nearest-neighbor hopping with respect to the nearest-neighbor hopping.

There are several unresolved issues in this work. First of all, why only the optimal doping of LCCO shifted to $n \sim 0.11$ while that of other electron-doped cuprates stay around 0.15 is not quite clear. There may be a combination of factors related to the largest $La^{3+}$ ionic radius, like the increased $-t'/t$, the suppressed AF exchange interaction $J$, and that the different radius of $La^{3+}$ influences the amount of apical oxygen ions. Secondly, the phase diagram of electron-doped LCCO and hole-doped LSCO show similar characteristics, which raises a possibility for absence of electron-hole asymmetry. But how to lead to the clarification of the electron-hole symmetry/asymmetry between hole-doped and electron-doped cuprates remains a mystery [1,3–6,13,26,27]. Thirdly, LCCO samples are grown on conducting Nb-SrTiO$_3$ (STO) substrates and successional annealing processes are necessary for ARPES measurements, so it is difficult to obtain $T_c$ after each procedure. It will be of significance to conduct transport measurements at samples with the same annealing procedures or on combinatorial films [28], so as to draw an accurate phase diagram based on electron doping level. Nevertheless, since the underdoped single-phase $T'$-LCCO films are difficult to be stabilized, this approach opens up a new avenue in understanding this material [29,30]. There needs more experimental and theoretical works for understanding the above issues.

In summary, using the elaborate two-step ozone/vacuum annealing method, we performed systematic ARPES measurements on electron-doped $T'$-LCCO for the first time. It was observed that the insulating behavior is absent at $n \sim 0.05$, probably due to the higher efficiency of removing apical oxygen ions by the annealing process. The Fermi surface reconstruction disappears at $n \sim 0.11$, indicating that the antiferromagnetism becomes weaker in the annealed LCCO films. By comparing with other electron-doped systems, we find that $-t'/t$ increases as increasing $Ln^{3+}$ ionic radius, which further suggests that the suppressed AF in LCCO may be contributed to the enhanced next-nearest-neighbor Cu hopping.

**Acknowledgements**
We thank K. Jiang and Y. G. Zhong for valuable discussions. This work was supported by the grants from the Natural Science Foundation of China (No.11888101, No.11227903), the Ministry of Science and Technology of China (2016YFA0401000), the Chinese Academy of Sciences (XDB07000000).

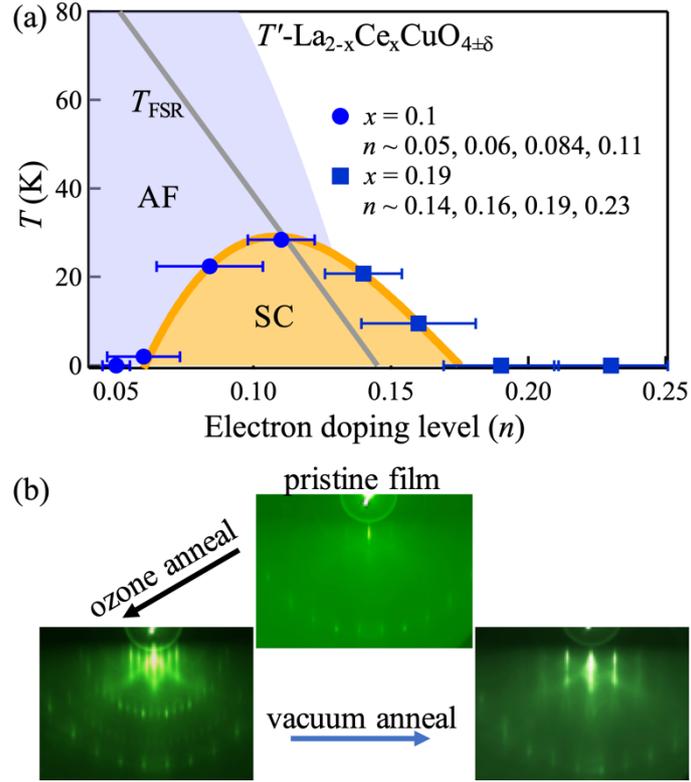

FIG. 1. (a) Phase diagram of $T'$-La$_{2-x}$Ce$_x$CuO$_{4\pm\delta}$, the superconducting dome and antiferromagnetic region are referenced from resistivity and angular magnetoresistance data, respectively [8,15]. The gray Fermi surface reconstruction (FSR) line is based on Hall measurements [16]. Blue marks represent the doping levels estimated from the FS volume of films by ozone/vacuum annealing, with the Ce concentrations $x$ corresponding to 0.1 (dots) and 0.19 (squares). Error bars are estimated from the uncertainty of determining the nodal $k_F$. (b) RHEED patterns record the process of refreshing the film surface and tuning the doping, which are from a pristine LCCO film exposed to air and after ozone/vacuum annealing.

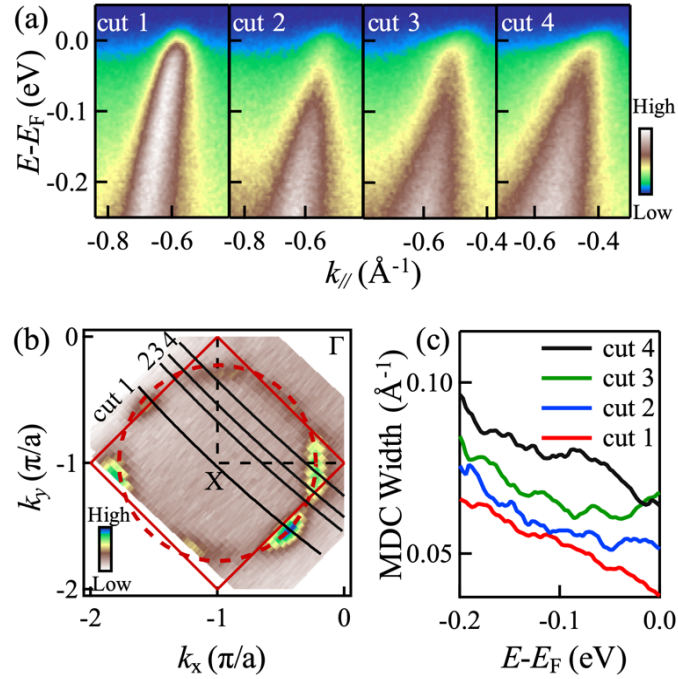

FIG. 2. (a) The representative band dispersions were performed at 10 K, which are referred to as node, hotspot, near hotspot and antinode, respectively, as indicated in (b). (b) Four-fold symmetrized Fermi surface measured at 30 K by integrating over $E_F \pm 15$ meV, the locus of the highest intensity in the FS map is fitted by the tight-binding model, which is shown as the red dashed curves. The estimated doping level is $n \sim 0.06$. The red solid lines are the AF Brillouin-zone boundaries. (c) The HWHM of MDCs for these cuts.

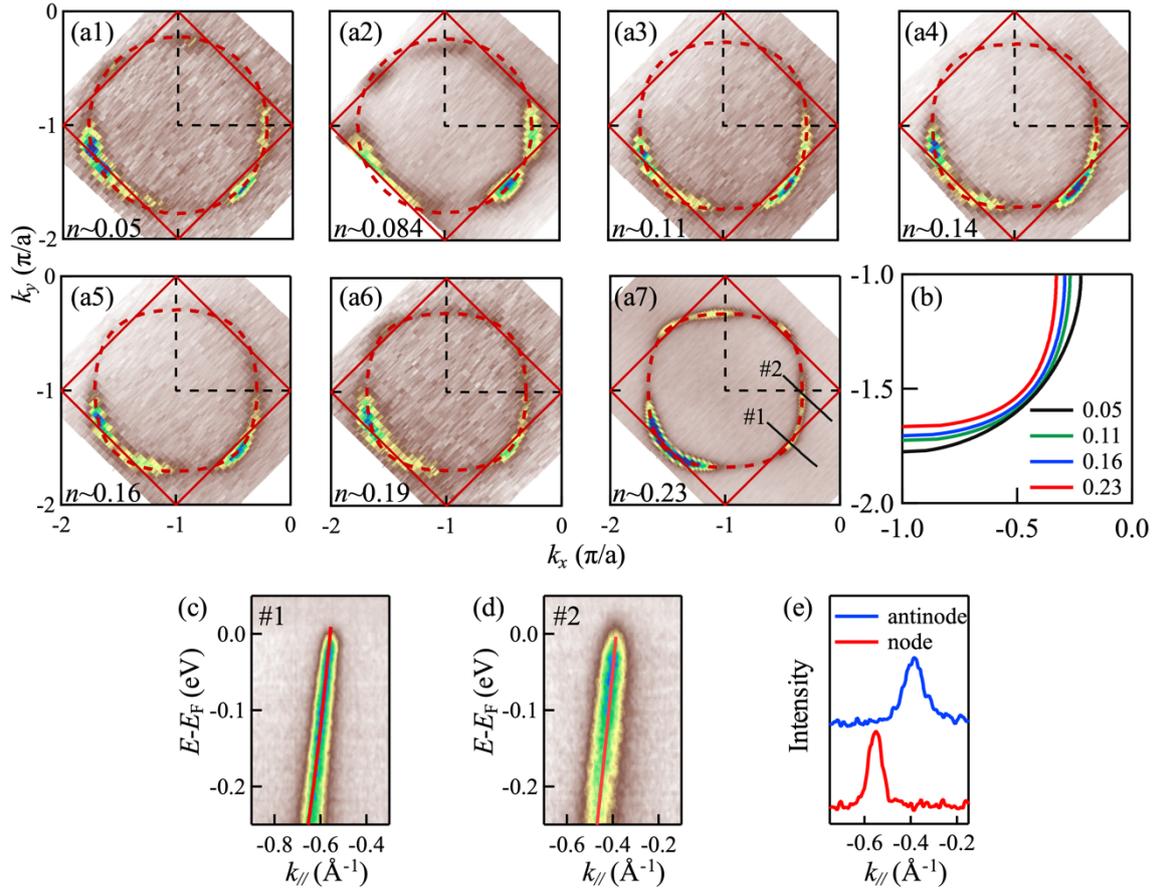

FIG. 3. (a1)–(a7) The FS evolution from n ~ 0.05 to 0.23. (b) The schematic FSs of films from highly underdoped to highly overdoped region. (c)–(d) Nodal and antinodal band dispersions were acquired at $n \sim 0.23$, which are indicated in (a7), the red lines are the tight–binding fitting results with $t = 0.3$. (e) MDCs at $E_F$ for (c) and (d).

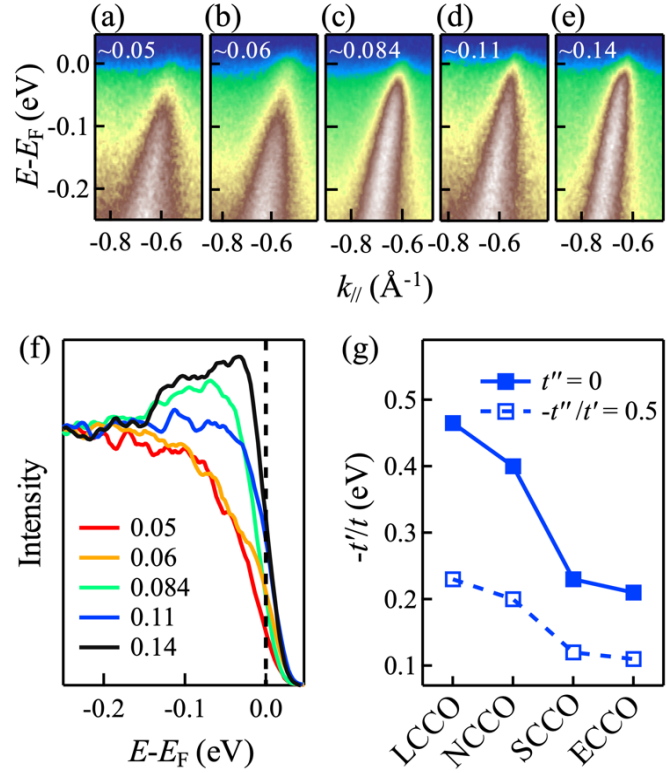

FIG. 4. (a) – (e) Doping evolution of band dispersions at the hotspot. Spectra at $n \sim 0.06, 0.084$ were measured at 10 K, others were measured at 30 K. The EDCs at $k_F$ are summarized in (b). (c) Comparison of $-t'/t$ for different systems LCCO, NCCO, SCCO and ECCO at their optimal doping. Solid and dashed lines reflect the assumption $t'' = 0$ and $-t''/t' = 0.5$, respectively.